\documentclass[11pt]{article}
\usepackage{psfig}
\usepackage{amsmath}
\usepackage{geometry}                
\usepackage{graphicx}
\usepackage{amssymb}
\usepackage{epstopdf}

\usepackage {natbib}
\begin{document}

\fontsize{11}{14.5pt}\selectfont

\begin{center}

{\small Technical Report No.\ 0606,
 Department of Statistics, University of Toronto}

\vspace*{0.7in}

{\LARGE\bf Gene Function Classification Using \\[7pt] Bayesian Models with
Hierarchy-Based Priors}\\[12pt]

 \parbox[t]{7.6cm} {
 \begin{center}
{\large Babak Shahbaba}\\[2pt]
 Dept.\ of Public Health Sciences, Biostatistics\\ 
 University of Toronto \\
 Toronto, Ontario, Canada \\
 \texttt{babak@stat.utoronto.ca}\\[10pt]
 \end{center}
}
\hfill
\parbox[t]{8.5cm} {
\begin{center}
{\large Radford M. Neal}\\[2pt]
 Dept.\ of Statistics and Dept.\ of Computer Science\\
 University of Toronto \\
 Toronto, Ontario, Canada \\
 \texttt{radford@stat.utoronto.ca}\\[10pt]
 \end{center}
 }

5 May 2006

\end{center}

\vspace*{8pt}

\noindent \textbf{Abstract.}  We investigate the application of
hierarchical classification schemes to the annotation of gene function
based on several characteristics of protein sequences including
phylogenic descriptors, sequence based attributes, and predicted
secondary structure. We discuss three Bayesian models and compare
their performance in terms of predictive accuracy. These models are
the ordinary multinomial logit (MNL) model, a hierarchical model based
on a set of nested MNL models, and a MNL model with a prior that
introduces correlations between the parameters for classes that are
nearby in the hierarchy. We also provide a new scheme for combining
different sources of information. We use these models to predict the
functional class of Open Reading Frames (ORFs) from the \emph{E. coli}
genome.  The results from all three models show substantial
improvement over previous methods, which were based on the C5
algorithm. The MNL model using a prior based on the hierarchy
outperforms both the non-hierarchical MNL model and the nested MNL
model. In contrast to previous attempts at combining these sources of
information, our approach results in a higher accuracy rate when
compared to models that use each data source alone. Together, these
results show that gene function can be predicted with higher accuracy
than previously achieved, using Bayesian models that incorporate
suitable prior information.

\section{\hspace*{-7pt}Introduction}\vspace*{-10pt}

Annotating genes with respect to the function of their proteins is essential for understanding the wealth of genomic information now available. A direct approach to identifying gene function is to eliminate or inhibit expression of a gene and observe any alteration in the phenotype. However, analysis of all genes for all possible functions is not possible at present. Statistical methods have therefore been employed for this purpose. One statistical approach attempts to predict the functional class of a gene based on similar sequences for which the function is known. The similarity measures used for this task are produced by computer algorithms that compare the sequence of interest against all other sequences with known function. The most commonly used algorithms are BLAST (\citealt{altschul97}) and FASTA (\citealt{pearson88}).

A problem with using such similarity measures is that a gene's function cannot be predicted when no homologous gene of known function exists. To improve the quality and coverage of prediction, other sources of information can be used. For example, \citet{king01} used a variety of protein sequence descriptors, such as residue frequency and the predicted secondary structure (the structure of hydrogen binding between different residues within a single polypeptide chain). \citet{derisi97}, \citet{eisen98} and \citet{brown00} used gene expression data, on the assumption that similarly expressed genes are likely to have similar function. \citet{marcotte99} recommended an alternative sequence-based approach, called the ``Rosetta stone'' method, which regards two genes as similar if they are together in another genome. \citet{deng03} predict the function of genes from their network of physical interactions. To address some of the problems associated with similarity-based methods, such as their non-robustness to variable mutation rates (\citealt{eisenja98}; \citealt{rost02}), annotation of protein sequences using phylogenetic information has been suggested by some authors (e.g., \citealt{eisen98}; \citealt{sjolander04}; \citealt{engelhardt05}). In this approach, the evolutionary history of a specific protein, captured by a phylogenetic tree, is used for annotating that protein (\citealt{eisen98}). 

The above-mentioned sources of data can be used separately, or as proposed by several authors (e.g.,  \citealt{king01}; \citealt{pavlidis01}; \citealt{deng04}), they can be combined within a predictive model. A variety of statistical and machine learning techniques for making such predictions have been used in functional genomics. These include neighbourhood-count methods (\citealt{schwikowski00}), support vector machines (\citealt{brown00}), decision tree models (\citealt{king01}), and Markov random fields (\citealt{deng03}). A common feature of these models is that they treat classes as unrelated entities without any specific structure. 

The assumption of unrelated classes is not always realistic. As argued by \citet{rison00}, in order to understand the overall mechanism of the whole genome, the functional classes of genes need to be organized according to the biological processes they perform. For this purpose, many functional classification schemes have been proposed for gene products. The first such scheme was recommended by \citet{riley93} to catalogue the proteins of \emph{Escherichia coli}. Since then, there have been many attempts to provide a standardized functional annotation scheme with terms that are not limited to certain types of proteins or to specific species. These schemes usually have a hierarchical structure, which starts with very general classes and becomes more specific in lower levels of the hierarchy. In some classification hierarchies, such as the Enzyme Commission (EC) scheme (\citealt{iubmb92}), levels have semantic values (\citealt{rison00}). For example, the first level of the EC scheme represents the major activities of enzyme like ``transferaces'' or ``hydrolases''. In some other schemes, like the ones considered here, the levels do not have any uniform meaning. Instead, each division is specific to the parent nodes. For instance, if the parent includes ``metabolism'' functions, the children nodes could be the metabolism of ``large'' or ``small'' molecules. \citet{rison00} surveyed a number of these structures and compared them with respect to their resolution (total number of function nodes), depth (potential of the scheme for division into subsets) and breadth (number of nodes at the top level).   

All these hierarchies provide additional information that can be incorporated into the classification model. For example, \citet{king01} attempted to use the additional information from the hierarchical structure of functional classes in \emph{E. coli} by simply using different decision tree models for each level of the hierarchy. \citet{clare03} expanded this approach by modifying the original decision tree model so that the assignment of a functional class to a node in the decision tree implied membership of all its parent classes. They evaluated this method based on \emph{Saccharomyces cerevisiae} data and found that the modified version is sometimes better than the non-hierarchical model and sometimes worse.

In a previous paper (\citealt{shahbaba05}), we introduced an alternative Bayesian framework for modelling hierarchical classes. This method uses a Bayesian form of the multinomial logit model, with a prior that introduces correlations between the parameters for classes that are nearby in the tree. We also discussed an alternative hierarchical model that uses the hierarchy to define a set of nested multinomial logit models. In this paper, we apply these methods to the gene function classification problem. 

The rest of this paper is organized as follows. In the next section, we explain our general method using a simple hierarchy for illustration. In section 3, we describe the dataset we used to test our approach. The details of the models we used and their implementation are in sections 4 and 5. The results of our analysis are presented in section 6. Section 7 is devoted to discussion, limitations of the proposed method, and future directions.

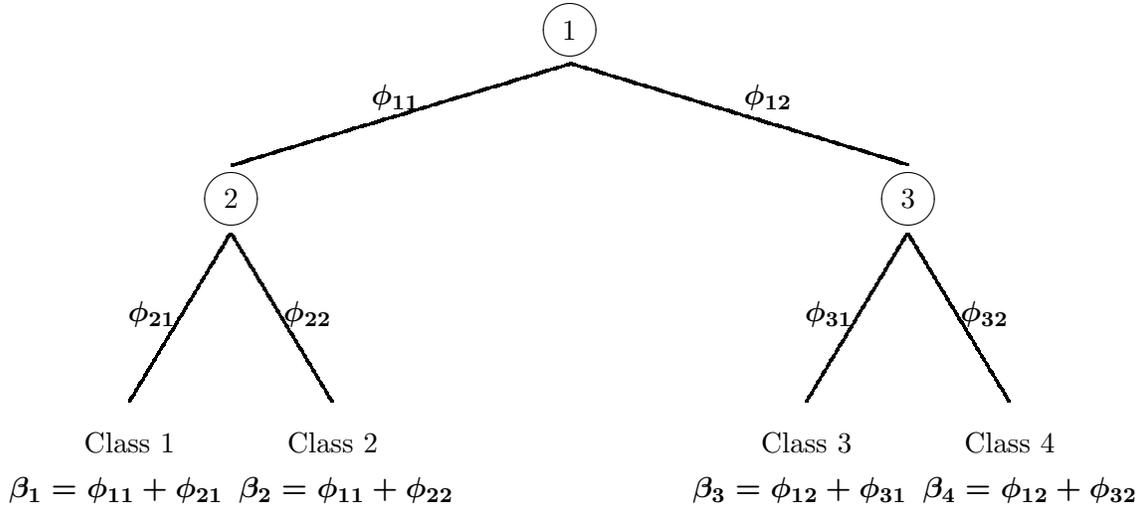
\begin{figure*} \label{tree}
\unitlength .9mm
\hspace*{-30pt}\begin{picture}(147,75)(-25,0)
\put(55,60){\makebox(0,0)[cc]{$\boldsymbol{\phi_{11}}$ }}

\put(110,60){\makebox(0,0)[cc]{$\boldsymbol{\phi_{12}}$ }}

\put(31.5,36){\makebox(0,0)[cc]{}}

\put(19,28){\makebox(0,0)[cc]{$\boldsymbol{\phi_{21}}$ }}

\put(42,28){\makebox(0,0)[cc]{$\boldsymbol{\phi_{22}}$ }}

\put(119,28){\makebox(0,0)[cc]{$\boldsymbol{\phi_{31}}$ }}

\put(142,28){\makebox(0,0)[cc]{$\boldsymbol{\phi_{32}}$ }}

\put(15,9){\makebox(0,0)[cc]{Class 1}}

\put(45,9){\makebox(0,0)[cc]{Class 2}}

\put(115,9){\makebox(0,0)[cc]{Class 3}}

\put(145,9){\makebox(0,0)[cc]{Class 4}}

\put(13,2){\makebox(0,0)[cc]{$\boldsymbol{\beta_1 = \phi_{11} + \phi_{21} }$}}

\put(47,2){\makebox(0,0)[cc]{$\boldsymbol{\beta_2 = \phi_{11} + \phi_{22} }$}}

\put(114,2){\makebox(0,0)[cc]{$\boldsymbol{\beta_3 = \phi_{12} + \phi_{31} }$}}

\put(148,2){\makebox(0,0)[cc]{$\boldsymbol{\beta_4 = \phi_{12} + \phi_{32} }$}}

\linethickness{0.3mm}
\multiput(15,15)(0.12,0.2){125}{\line(0,1){0.2}}
\linethickness{0.3mm}
\multiput(30,40)(0.12,-0.2){125}{\line(0,-1){0.2}}
\linethickness{0.3mm}
\linethickness{0.3mm}
\multiput(115,15)(0.12,0.2){125}{\line(0,1){0.2}}
\linethickness{0.3mm}
\multiput(130,40)(0.12,-0.2){125}{\line(0,-1){0.2}}
\linethickness{0.3mm}
\put(80,70){\circle{8.33}}

\put(80,70){\makebox(0,0)[cc]{1}}

\put(105,75){\makebox(0,0)[cc]{}}

\linethickness{0.3mm}
\multiput(30,50)(0.4,0.12){125}{\line(1,0){0.4}}
\linethickness{0.3mm}
\multiput(80,65)(0.4,-0.12){125}{\line(1,0){0.4}}
\linethickness{0.3mm}
\put(30,45){\circle{8.33}}

\linethickness{0.3mm}
\put(130,45){\circle{8.33}}

\put(30,45){\makebox(0,0)[cc]{2}}

\put(130,45){\makebox(0,0)[cc]{3}}

\end{picture}

\caption{The corMNL model for a simple hierarchy. The coefficient parameter for each class is a sum of parameters at different levels of the hierarchy }
\label{simpleHierarchy}
\end{figure*}

\section{\hspace*{-7pt}Methodology}\vspace*{-10pt}

When classes in a classification problem are unrelated, a simple multinomial logit (MNL) model may be appropriate. Consider a classification problem in which we have observed data for $n$ cases, ($x^{(1)},y^{(1)}$), ...,($x^{(n)},y^{(n)}$), where $x^{(i)} = x_{1}^{(i)}, ..., x_{p}^{(i)}$ is the vector of $p$ covariates (features) for case $i$, and $y^{(i)}$ is the associated class. The following is the MNL model, which is also known as ``softmax'' in the machine learning literature: 
\begin{eqnarray*} \label{mnl}
P(y=j|x, \boldsymbol{\alpha}, \boldsymbol{\beta}) & = &\frac{\exp(\alpha_j + x\boldsymbol{ \beta}_j)}{\sum_{j'=1}^{c} \exp(\alpha_{j'}+x\boldsymbol{ \beta}_{j'})}
\end{eqnarray*}
Here, $c$ is the number of classes. For each class, $j$, there is an intercept $\alpha_j$ and a vector of $p$ unknown parameters $\boldsymbol \beta_j$. The inner product of these parameters with the covariate vector is shown as $x\boldsymbol{\beta}_j$. The entire set of regression coefficients $\boldsymbol{\beta = \beta_1, ..., \beta_c}$ can be presented as a $p \times c$ matrix. This representation is redundant, since one of the $\boldsymbol{\beta}_j$'s can be set to zero without changing the set of relationships expressible with the model. In methods based on maximum likelihood estimation, it is common to set either $\beta_1$ or $\beta_c$ to zero. However, in a Bayesian framework, the redundant representation is preferred, since removing this redundancy would make it difficult to specify a prior that treats all classes symmetrically. Priors such as the following are typically used:
\begin{eqnarray*}
\alpha_{j} | \eta   & \sim & N(0, \eta^{2}) \\
\beta_{jl} | \tau  & \sim & N({0}, \tau^{2}) \\
log(\eta) & \sim & N(v, V^2)  \\
log(\tau) & \sim & N(w, W^2)
\end{eqnarray*}  
where $j = 1, ..., c$ and $l = 1, ..., p$. Here and later, independence is assumed unless conditioning variables are shown.

For problems such as gene function classification, the assumption of unrelated classes does not always hold. In many cases, classes have a hierarchical structure. The importance of using the hierarchy in classifiers has been emphasized by many authors (e.g., \citealt{sattath77}; \citealt{fox97}; \citealt{koller97}). One approach for modelling hierarchical classes is to decompose the classification model into nested models (e.g., logistic or MNL). For hierarchical classification problems with simple binary partitions, \citet{fox97} suggested using successive logistic models for each binary class. In Figure \ref{simpleHierarchy} below, for example, these partitions are \{12, 34\}, \{1, 2\}, and \{3, 4\}. The resulting nested binary models are statistically independent. The likelihood can therefore be written as the product of the likelihoods for each of the binary models. For example, in Figure \ref{simpleHierarchy} we have
\begin{eqnarray*}
P(y=1| x) & = & P(y \in \{1, 2 \} | x) \times P(y \in \{1\}| y \in \{1, 2\}, x)
\end{eqnarray*}
Restriction to binary models is unnecessary. At each level, classes can be divided into more than two subsets and MNL can be used instead of logistic regression. We refer to Bayesian models in which the tree structure is used to define a set of nested MNL models as treeMNL. Consider a parent node, $m$, with $c_m$ child nodes, representing sets of classes $S_k$, for $k = 1, ..., c_m$. The portion of the treeMNL model for this node has the form:
\begin{eqnarray*}
P(y \in S_k|x, \boldsymbol{\alpha}_m, \boldsymbol{\beta}_m) & = &\frac{\exp(\alpha_{mk} + x\boldsymbol{ \beta}_{mk})}{\sum_{k'=1}^{c_m} \exp(\alpha_{m{k'}}+x\boldsymbol{ \beta}_{m{k'}})}\\
\alpha_{mk} | \eta_{m} \  & \sim & N(0, \eta_{m}^{2}) \\
\beta_{mkl} | \tau_{m} & \sim & N({0}, \tau_{m}^{2}) \\
log(\eta_{m}) & \sim & N(v_m, V_{m}^{2})  \\
log(\tau_{m})& \sim & N(w_m, W_{m}^{2})
\end{eqnarray*}  
We calculate the probability of each end node, $j$, by multiplying the probabilities of all intermediate nodes leading to $j$.

In contrast to this treeMNL model, \citet{mitchell98} showed that the hierarchical naive Bayes classifier is equivalent to the standard non-hierarchical classifier when probabilities are estimated by maximum likelihood (ML). To improve the hierarchical naive Bayes model, \citet{mccallum98} suggested smoothing the parameter estimate for each end node by shrinking its ML estimate towards the estimates for all its ancestors in the hierarchy. More recently, new hierarchical classification models based on Support Vector Machines (SVM) have been proposed (\citealt{dumais00}; \citealt{dekel04}; \citealt{cai04}; \citealt{tsochantaridis04}; \citealt{cesa06}). An important aspect of these models is the use of a modified loss function which reflects the taxonomy of classes.

For modelling hierarchical classes, we introduced a new method which has a MNL form with a prior that introduces correlations between the parameters of nearby classes (\citealt{shahbaba05}). Our model includes a vector of parameters for each branch in the hierarchy. We assign objects to one of the end nodes using a MNL model whose regression coefficients for class $j$ are represented by the sum of the parameters for all the branches leading to that class. Sharing of common parameters (from common branches) introduces prior correlations between the parameters of nearby classes in the hierarchy. This way, we can better handle situations in which these classes are hard to distinguish. Our simulation results show that when the hierarchy actually provides information about how distinguishable classes are, our model, which we call corMNL, outperforms both the non-hierarchical MNL model and the nested treeMNL model (\citealt{shahbaba05}). When an inappropriate hierarchy is used, the penalty for corMNL is significantly less than for the alternative treeMNL model. 

Consider Figure \ref{simpleHierarchy}, which shows a hierarchical classification problem with four classes. Parameter vectors denoted as $\boldsymbol{\phi_{11}$ and $\phi_{12}}$ are associated with branches in the first level, and $\boldsymbol{\phi_{21}}$, $\boldsymbol{\phi_{22}}$, $\boldsymbol{\phi_{31}}$ and $\boldsymbol{\phi_{32}}$ with branches in the second level. We assign objects to one of the end nodes using a MNL model whose regression coefficients for a class are represented by the sum of parameters on all the branches leading to that class. In Figure \ref{simpleHierarchy}, these coefficients are $\boldsymbol{\beta_{1} = \phi _{11}+\phi _{21} }$, $\boldsymbol{\beta_{2} = \phi_{11}+\phi _{22} }$, $\boldsymbol{\beta_{3} = \phi _{12}+\phi _{31} }$ and $\boldsymbol{\beta_{4} = \phi _{12}+\phi _{32} }$ for classes $1, 2, 3$ and $4$ respectively. Sharing the common terms, $\boldsymbol{\phi_{11}}$ and $\boldsymbol{\phi_{12}}$, introduces prior correlation between the parameters of nearby classes in the hierarchy. Note that the intercept parameters, $\alpha_j$, are not treated hierarchically.

In our model, $\boldsymbol{\phi}$'s are vectors with the same size as $\boldsymbol{\beta}$'s. We assume that, conditional on higher level hyperparameters, all the components of the $\boldsymbol\phi$'s are independent, and have normal prior distributions with zero mean. The variances of these components are regarded as hyperparameters, which control the magnitudes of coefficients. When a part of the hierarchy is irrelevant, we hope the posterior distribution of its corresponding hyperparameter will be concentrated near zero, so that the parameters it controls will also be close to zero. In detail, the simplest form of prior for a corMNL model is as follows:
\begin{eqnarray*}
\alpha_j | \eta \  & \sim & N(0, \eta^{2})  \\
{\phi}_{mkl} | \tau_{m} & \sim & N({0}, \tau_{m}^{2}) \\
log(\eta) & \sim & N(v, V^2)  \\
log(\tau_{m}) & \sim & N(w_m, W_{m}^{2})
\end{eqnarray*} 
Here, $\phi_{mkl}$ refers to the parameter related to covariate $x_l$ and branch $k$ of node $m$. For gene function classification, we used a more elaborate prior, discussed in section 4.

\section{\hspace*{-7pt}Data}\vspace*{-10pt}

We used our model to predict the functional class of Open Reading Frames (ORFs) from the \emph{E. coli} genome. \emph{E. coli} is a good organism for testing our method since many of its gene functions have been identified through direct experiments. We used the pre-processed data provided by \citet{king01}. This dataset contains 4289 ORFs identified by \citet{blattner97}. Only 2122 of these ORFs, for which the function is known, are used in our analysis. The functional hierarchy for these proteins is provided by \citet{riley96}. This hierarchy has three levels with the most general classes at level 1 and the most specific classes at level 3. For example, lipoate-protein ligase A (lplA) belongs to class `Macromolecule metabolism' at level 1, to class `Macromolecule synthesis, modification' at level 2, and to class `Lipoprotein' at level 3. After excluding categories 0 and 7 at level 1, the data we used had 6 level 1 categories, 20 level 2 categories, and 146 level 3 categories. 

It is worthwhile mentioning that since 2001 the function of many new genes have been determined by direct experiment. However, we use the same dataset as \citet{king01}, with the same split of data into the training set (1410 ORFs) and test set (712 ORFs), in order to produce comparable results. \citet{king01} further divided the training set into two subsets and used one subset as validation data to select a subset of rules from those produced by the C5 algorithm based on the other part of the training set. Our Bayesian methods do not require a validation set, so we did not subdivided the training set.  

The covariates are based on three different sources of information: phylogenic descriptors, sequence based attributes, and predicted secondary structure. Following \citet{king01}, we refer to these three sources of data as SIM, SEQ and STR respectively. Attributes in SEQ are essentially based on the composition of singlets and pairs of residues in a sequence. There are 933 such attributes. Information in SIM and STR is derived based on a PSI-BLAST (position-specific iterative BLAST) search with parameters $e=10$, $h = 0.0005$, $j=20$ from NRProt 05/10/99 database. \citet{king01} used the Inductive Logic Programming (ILP) algorithm known as Warmr (\citealt{dehaspe98}) to produce binary attributes based on the identified frequent patterns (1 if the pattern is present and 0 otherwise) in SIM and STR data. There are 13799 such attributes generated for SIM and 18342 attributes for STR. 

\section{\hspace*{-7pt}Models}\vspace*{-10pt}

We first used our models to predict gene function using each data source (SIM, STR and SEQ) separately. Since the numbers of covariates in these datasets are large, we applied principal Component Analysis (PCA). Prior to applying PCA, the variables were centred to have mean zero, but they were not rescaled to have variance one. We selected the first $p$ components with the highest eigenvalues. The cutt-off, $p$, was set based on the plot of eigenvalues against PCs (i.e., the scree plot). Since there was not a clear cut-off point at which the magnitude of eigenvalues drops sharply, the plots could only help us to narrow down the appropriate values for $p$. We decided to choose a value at the upper end of the range suggested by the scree plot. We selected 100 components from SEQ, 100 components from STR, and 150 components from SIM.  

Principal components are derived solely based on the input space and do not necessarily provide the best set of variables for predicting the response variable. In order to find the relevant variables (among the principal components) for the classification task, we use the Automatic Relevance Determination (ARD) method suggested by \citet{neal96}. ARD employs a hierarchical prior to determine how relevant each covariate is to classification. In the MNL model, for example, one hyperparameter, $\sigma_l$, is used to control the variance of all coefficients, $\beta_{jl}$ ($j=1, ..., c$), for covariate $x_l$. If a covariate is irrelevant, its hyperparameter will tend to be small, forcing the coefficients for that covariate to be near zero. We also use a set of hyperparameters, $\tau_j$, to control the magnitude of the $\beta$'s for each class. We use a third hypeparameter, $\xi$, to control the overall magnitude of all $\beta$'s. This way, $\sigma_l$ controls the relevance of covariate $x_l$ compared to other covariates, $\tau_j$, controls the usefulness of covariates in identifying class $j$, and $\xi$ controls the overall usefulness of all covariates in separating all classes. The standard deviation of $\beta_{jl}$ is therefore equal to $\xi \tau_j \sigma_{l}$. 

For the MNL model we used the following priors:
\begin{eqnarray*}
\alpha_j | \eta & \sim & N(0, \eta^{2}) \\
\beta_{jl} | \xi, \sigma_l, \tau & \sim & N({0}, \xi^{2} \tau_{j}^{2}\sigma_{l}^{2})  \\
log(\eta) & \sim & N(0, 1) \\
log(\xi) & \sim & N(-3, 2^2) \\
log(\tau_{j}) & \sim & N(-1, 0.5^2)  \\
log(\sigma_{l}) & \sim & N(0, 0.3^2) 
\end{eqnarray*} 
Since the task of variable selection is mainly performed through PCA, the ARD hyperparameters, $\sigma$'s, are given priors with fairly small standard deviation. The priors for $\tau$'s are set such that both small values (i.e., close to zero) and large values (i.e., close to 1) are possible. The overall scale of these hyperparameters is controlled by $\xi$, which has a broader prior. Note that since these hyperparameters are used only in the combination $\xi \tau_j \sigma_l $, only the sum of the means for $log(\xi)$, $log(\tau_j)$, and $log(\sigma_l)$ really matters. 
   
Similar priors are used for the parameters of treeMNL and corMNL. For these two models, we again used one hyperparamter, $\sigma_{l}$, to control all parameters ($\beta$'s in treeMNL, $\phi$'s in corMNL) related to covariate $x_l$. We also used one scale parameter $\tau_{k}$ for all parameters related to branch $k$ of the hierarchy. The overall scale of all parameters is controlled by one hyperparameter $\xi$. 

This setting of priors is different from what we used in a previous paper (\citealt{shahbaba05}), where we used one hyperparameter to control all the coefficients (regardless of their corresponding class) in the MNL model, and we used one hyperparameter to control the parameters of all the branches that share the same node in treeMNL and corMNL. The scheme used in this paper provides an additional flexibility to control $\beta$'s. In this paper, the hyperparameters are given log-normal distributions instead of the gamma distributions used in \citet{shahbaba05}. Using gamma priors has the advantage of conjugacy and, therefore, easier MCMC sampling. However, we prefer log-normal distribution since they are more convenient for formalizing our prior beliefs. 

\section{\hspace*{-7pt}Implementation}\vspace*{-10pt}

These models are implemented using Markov chain Monte Carlo (\citealt{neal93}). We use Hamiltonian dynamics (\citealt{neal93}) for sampling from the posterior distribution of coefficients (with hyperparameters temporarily fixed). The number of leapfrog steps was set to $50$. The stepsizes were set dynamically at each iteration, based on the current values of the hyperparameters (\citealt{neal96}). In the MNL and corMNL models, new values are proposed for all regression parameters simultaneously. Nested MNL models in treeMNL are updated separately since they are regarded as independent models. The coefficient parameters within each nested model, however, are updated at the same time. 

We use single-variable slice sampling (\citealt{neal03}) to sample from the posterior distribution of hyperparameters. At each iteration, we use the ``stepping out'' procedure to find the interval around the current point and the ``shrinkage'' procedure for sampling from the interval. The initial values of the ARD hyperprameters, $\sigma$'s, were set to the inverse of the standard deviation of their corresponding covariates. The initial values of $\tau$'s and $\xi$ were set to 1. 

Convergence of the Markov chain simulations was assessed from trace plots of hyperparameters. We ran each chain for 5000 iterations, of which the first 1000 were discarded. Simulating the Markov chain for 10 iterations took about 2 minutes for MNL, 1 minute for treeMNL, and 3 minutes for corMNL, using a MATLAB implementation on an UltraSPARC III machine. 

\begin{table*}
\begin{center}
\begin{tabular}{l || c|c|c|| c|c|c|| c|c|c||}
Accuracy (\%) & \multicolumn{3}{c||}{ SEQ} & \multicolumn{3}{c||}{ STR} & \multicolumn{3}{c||}{ SIM}\\
 & Level 1 & Level 2 & Level 3 & Level 1 & Level 2 & Level 3 & Level 1 & Level 2 & Level 3 \\
\hline&&&&&&&&&\\[-12pt]\hline
Baseline & 42.56 & 21.21 & 8.15 & 42.56 & 21.21 & 8.15 & 42.56 & 21.21 & 8.15 \\
\hline
MNL &  60.25 & 33.99 & 20.93  &  50.98 & 25.14 & 15.87   &  69.10 & 45.79 & 30.76 \\
\hline
treeMNL & 59.27 & 34.13 & 18.26  &    52.67 & 27.39 & 16.29     & 67.70 & 45.93 & 30.34 \\
\hline
corMNL & \textbf{61.10} & \textbf{35.96} & \textbf{21.21} & \textbf{52.81} & \textbf{27.95} & \textbf{16.71} & \textbf{70.51} & \textbf{47.19} & \textbf{30.90} 
\end{tabular}
\end{center}
\caption{Comparison of models based on their predictive accuracy (\%) using each data source separately.}
\label{tab:sepRes}
\vspace*{20pt}

\begin{center}
\begin{tabular}{l || c|c|c|| c|c|c|| c|c|c||}
Accuracy (\%) & \multicolumn{3}{c||}{ SEQ} & \multicolumn{3}{c||}{ STR} & \multicolumn{3}{c||}{ SIM}\\
 & Level 1 & Level 2 & Level 3 & Level 1 & Level 2 & Level 3 & Level 1 & Level 2 & Level 3 \\
& (20)  &   (18)  &  (4)   & (10)  &  (1)   &  (5)    &  (29)  &  (26)  &  (16) \\
\hline&&&&&&&&&\\[-12pt]\hline
C5 & 64 & 63 & 41 & 59 & 44 & 17 & 75 & 74 & 69 \\
\hline
MNL &  81 & 79 & 88  &  \textbf{83} & \textbf{100} & 67  &  96 & \textbf{90} & \textbf{84} \\
\hline
treeMNL & 81 & 76 & 70  &    70 & 86 & 69  & 95 & 87 & 84 \\
\hline
corMNL & \textbf{84} & \textbf{82} & \textbf{89} & \textbf{83} & \textbf{100} & \textbf{73} & \textbf{97} & \textbf{90} &{82} 
\end{tabular}
\end{center}
\caption{Comparison of models based on their predictive accuracy (\%) for specific coverage (provided in parenthesis). The C5 results and the coverage values are from \cite{king01}.}
\label{tab:kingRes}
\vspace*{20pt}

\begin{center}
\begin{tabular}{l || c|c|c|| c|c|c|| c|c|c||}
Accuracy (\%) & \multicolumn{3}{c||}{ SIM only } & \multicolumn{3}{c||}{ Combined dataset } & \multicolumn{3}{c||}{ Combined dataset } \\
& \multicolumn{3}{c||}{}&  \multicolumn{3}{c||}{ single scale parameter } &  \multicolumn{3}{c||}{ separate scale parameters } \\
& \multicolumn{3}{c||}{}&  \multicolumn{3}{c||}{ } &  \multicolumn{3}{c||}{ } \\ 
  & Level 1 & Level 2 & Level 3 & Level 1 & Level 2 & Level 3 & Level 1 & Level 2 & Level 3 \\
\hline&&&&&&&&&\\[-12pt]\hline
MNL &  69.10 & 45.79 & 30.76 & 69.66 & 48.88 & 32.02  &  70.65 & \textbf{49.16} & 33.71 \\
\hline
treeMNL & 67.70 & 45.93 & 30.34 & 68.26 & 46.63 & 30.34  & 68.82 & 46.63 & 31.74  \\
\hline
corMNL &  \textbf{70.51} & \textbf{47.19} & \textbf{30.90} & \textbf{71.49} & \textbf{49.30} & \textbf{32.87}  & \textbf{72.75} & \textbf{49.16} & \textbf{34.41} 
\end{tabular}
\caption{Accuracy (\%) of models on the combined dataset with and without separate scale parameters. Results based on using SIM alone are provided for comparison.}
\label{tab:strat}
\end{center}
\end{table*}

\section{\hspace*{-7pt}Results}\vspace*{-10pt}

Table \ref{tab:sepRes} compares the three models with respect to their accuracy of prediction at each level of the hierarchy. In this table, level 1 corresponds to the top level of the hierarchy, while level 3 refers to the most detailed classes (i.e., the end nodes). For level 3, we use a simple 0/1 loss function and minimize the expected loss by assigning each test case to the end node with the highest posterior predictive probability. We could use the same predictions for measuring the accuracy at levels 1 and 2, however, to improve accuracy, we instead make predictions based on the total posterior predictive probability of nodes at levels 1 and level 2.

To provide a baseline for interpreting the results, for each task we present the performance of a model that ignores the covariates and simply assigns genes to the most common category at the given level in the training set. 

As we can see in Table \ref{tab:sepRes}, corMNL outperforms all other models. For the SEQ dataset, MNL performs better than treeMNL. Compared to MNL, the corMNL model achieves a slightly better accuracy at level 3 and more marked improvements at level 1 and level 2. For the STR dataset, both hierarchical models (i.e., treeMNL and corMNL) outperform the non-hierarchical MNL. For this dataset, corMNL has a slightly better performance than treeMNL. For the SIM dataset, the advantage of using the corMNL model is more apparent in the first and second levels.

\citet{king01} used a decision tree model based on the C5 algorithm for analysing these datasets. They selected sets of rules that had an accuracy of at least 50\% with the coverage of at least two correct examples in the validation set. In Table \ref{tab:kingRes}, we compare the accuracy of our models to those of \citet{king01}. In order to make the results comparable, we used the same coverage values as they used. Coverage is defined as the percentage of test cases for which we make a confident prediction. In a decision tree model, these test cases can be chosen by selecting rules that lead to a specific class with high confidence. For our models, we base confidence on posterior predictive probability. We rank the test set based on these probabilities and for a coverage of $g$, we select the top $g$ percent. In Table \ref{tab:kingRes}, the coverage values are given in parenthesis. All three models discussed here outperform the decision tree model. Overall, corMNL has better performance than MNL and treeMNL.

\cite{king01} attempted to improve predictive accuracy by combing the three datasets (SEQ, STR and SIM). Although one would expect to obtain better predictions by combining several sources of information, their results showed no additional benefit compared to using the SIM dataset alone. We also tried combining datasets in order to obtain better results. Initially, we used the principal components which we found individually for each dataset, and kept the number of covariates contributed from each data source the same as before (i.e., 100 covariates from SEQ, 100 covariates from STR, and 150 covariates from SIM). Principal components from each dataset were scaled so that the standard deviation of the first principal component was 1. We did this to make the scale of variables from different data sources comparable while preserving the relative importance of principal components within a dataset. 

Using the combined dataset, all our models provided better predictions, although the improvement was only marginal for some predictions. We speculated that some of the variables (i.e., PCs) may become redundant after combining the data. That is, some variables are providing the same information. In general, one may obtain better results by removing redundancy and reducing the number of variables. To examine this idea, we kept the number of principal components from SIM as before (i.e., 150) but only used the first 25 principal components from SEQ and STR. The total number of covariates was therefore 200. Reducing the number of covariates from SEQ and STR may also prevent them from overwhelming the covariates from SIM, which is the most useful single source. This strategy led to even higher accuracy rates compared to when we used the SIM dataset alone. The results are shown in Table \ref{tab:strat} (middle section). It is worth noting that using 25 principal components results in a lower performance (i.e., lower accuracy rate) when SEQ and STR are used individually in the models (results not shown).

To improve the models even further, we tried using separate scale parameters, $\xi$, for different sources of information. This way, we allow the coefficients from different data sources to have appropriately different variances in the model. This is additional to what ARD hyperparameters provide. As we can see in Table \ref{tab:strat} (right section), this strategy resulted in further improvements in the performance of the models. The posterior distribution of the $\xi$'s reflected the importance of each data source. In the MNL model, the posterior means of the three $\xi$'s were $2.90$, $0.86$ and $4.67$ for SEQ, STR and SIM respectively. The corresponding values in treeMNL were $2.39$, $0.85$ and $4.56$ and in corMNL $2.21$, $0.74$ and $3.63$.

We examined the idea of having separate $\xi$'s and larger numbers of covariates. We found that when we increased the number of principal components for SEQ and STR to 100, the accuracy of predictions mostly remained the same, though a few dropped slightly. 

In practice, we might be most interested in genes whose function can be predicted with high confidence. There is a trade-off between predictive accuracy and the percentage of the genes we select for prediction (i.e., coverage). Table \ref{tab:coverage} shows this trade-off for results on the test set from the corMNL model with three $\xi$ hyperparameters applied to the combined dataset. In this table, the accuracy rates for different coverage values are provided. As we can see, our model can almost perfectly classifies 10\% of the genes in the test set. 

The MATLAB programs for MNL, treeMNL and corMNL, as well as the combined dataset for \emph{E. coli}, are available online at http://www.utstat.utoronto.ca/$\sim$babak. 

\begin{table}
\begin{center}
\vspace*{10pt}
\begin{tabular}{l || c|c|c|c|c|c|}
Accuracy (\%)  & \multicolumn{6}{c|}{Coverage (\%)}\\
& \emph{5} & \emph{10} & \emph{20} & \emph{50} & \emph {90} & \emph{100}\\
\hline & & & & & & \\[-12pt] \hline 
Level 1 & 100 & 98 & 96 & 92 & 76 & 73\\
\hline
Level 2 & 100 & 98 & 96 & 71 & 53 & 49\\
\hline
Level 3 & 100& 97 & 80 & 52 & 36 & 34\\
\end{tabular}
\caption{Predictive accuracy (\%) for different coverage values (\%) of the corMNL model using all three sources with three $\xi$ hyperparameters.}
\label{tab:coverage}
\end{center}
\end{table}

\section{\hspace*{-7pt}Conclusions and Future Directions}\vspace*{-10pt}

In this paper, we investigated the use of hierarchical classification schemes to perform functional annotation of genes. If the hierarchy provides any information regarding the structure of gene function, we expected that this additional information would lead to better prediction of classes. To examine this idea, we compared three Bayesian models: a non-hierarchical MNL model, a hierarchical model based on nested MNL, referred to as treeMNL, and the corMNL model, which is a form of the multinomial logit model with a prior that introduces correlations between the parameters of nearby classes. We found corMNL provided better predictions in most cases. Moreover, we introduced a new approach for combining different sources of data. In this method, we use separate scale parameters for each data source in order to allow their corresponding coefficients have appropriately different variances. This approach provided better predictions compared to other methods.

While our emphasis in this paper was on the importance of using hierarchical schemes in gene classification, we also showed that even the non-hierarchical Bayesian MNL model outperforms previous methods that used the C5 algorithm. Overall, our results are encouraging for the prospect of accurate gene function annotation, and also illustrate the utility of a Bayesian approach with problem-specific priors. For our experiments, we used the pre-processed datasets provided by \citet{king01}, who used the Warmr (\citealt{dehaspe98}) algorithm to generate binary attributes. It is conceivable that the accuracy of predictions can be further improved by using other data processing methods. Similarly, it is possible that a method other than our use of PCA might be better for reducing dimensionally before doing classification.

In the \emph{E. coli} dataset, each ORF was assigned to only one function. This is not the case for some other hierarchies such as the MIPS functional classification defined for the genome of \emph{S. cerevisiae}, where an ORF may belong to more than one class. For such problems, one can modify the likelihood part of the models described here to handle this additional complexity. For example, if a gene can belong to several classes with equal probabilities, its contribution to the likelihood is proportional to the sum of probabilities of those classes.

The functional hierarchies considered here are simple tree-like structures. There are other hierarchical structures that are more complex than a tree. For example, one of the most commonly used gene annotation schemes, known as Gene Ontology (GO), is implemented as a directed acyclic graph (DAG). In this structure a node can have more than one parent. Our method, as it is, cannot be applied to these problems, but it should be possible to extend the idea of summing coefficients along the path to the class in order to allow for multiple paths. 

Our approach can also be generalized to problems where the relationship among classes can be described by more than one hierarchical structure. For these problems, different hyperparameters can be used for each hierarchy and predictions can be made by summing the parameters in branches from all these hierarchies.

\section*{Acknowledgements}\vspace*{-10pt}

We thank Ross D. King, Andreas Karwath, Amanda Clare and Luc Dehaspe for providing the processed \emph{E. coli} datasets. This research was supported by the Natural Sciences and Engineering Research Council of Canada. Radford Neal holds a Canada Research Chair in Statistics and Machine Learning.

\end{document}